\documentclass[reprint,superscriptaddress,floatfix,aps,prb]{revtex4-2}

\usepackage{header}

\begin{document}
\title{Thermometry for a Kagome Lattice Dipolar Rydberg Simulator}
\author{Erik Fitzner}
\affiliation{Institut f\"ur Theoretische Physik, Universit\"at T\"ubingen, Auf der Morgenstelle 14, 72076 T\"ubingen, Germany}
\author{Igor Lesanovsky}
\affiliation{Institut f\"ur Theoretische Physik, Universit\"at T\"ubingen, Auf der Morgenstelle 14, 72076 T\"ubingen, Germany}
\affiliation{School of Physics and Astronomy and Centre for the Mathematics and Theoretical Physics of Quantum nonequilibrium Systems, The University of Nottingham, Nottingham, NG7 2RD, United Kingdom}
\author{Bj\"orn Sbierski}
\affiliation{Institut f\"ur Theoretische Physik, Universit\"at T\"ubingen, Auf der Morgenstelle 14, 72076 T\"ubingen, Germany}
\date{\today}

\begin{abstract}
    We propose an accurate thermometry approach for Rydberg atom tweezer arrays combining data from correlation and local susceptibility measurements with a theoretical high-temperature expansion method for dynamic spin correlations. We apply our approach to a recent quantum simulation experiment [Bornet et al., arXiv 2602.14323] realizing an anti-ferromagnetic dipolar spin-$1/2$ XY model on the Kagome lattice. We obtain $T=0.55J$ and $S/N=0.67 \, \mathrm{ln}2$ for temperature and entropy respectively, showing that further experimental efforts are required to reach the putative quantum spin liquid regime.
\end{abstract}

\maketitle

\section{Introduction}
Quantum spin liquids (QSLs) are strongly correlated low-temperature states of quantum magnets that evade conventional symmetry breaking and are instead characterized by long-range entanglement and fractionalized excitations~\cite{Anderson1973,Kitaev2006,Balents2010,Savary2017}. However, despite substantial theoretical progress and the investigation of many candidate materials, unambiguous experimental evidence for QSLs remains elusive in solid-state systems, largely due to material-specific complications such as disorder~\cite{Norman2016,Wen2019,Savary2017}.
More recently, atomic quantum simulators have emerged as controlled platforms realizing relevant spin models~\cite{Lukin2003,Santos2004,Altman2021}. In particular, Rydberg atom tweezer arrays offer flexible frustrated geometries, well-controlled Hamiltonians $H$ and site-resolved measurement capabilities~\cite{browaeysManyBodyPhysicsIndividuallyControlled2020,steinertSpatiallyTunableSpin2023,bornet_enhancing_2024}. Consequently, enormous efforts have been directed to create highly entangled many-body states that are difficult to realize in conventional solid-state settings ~\cite{schollQuantumSimulation2D2021,ebadiQuantumPhasesMatter2021,Semeghini2021,chenContinuousSymmetryBreaking2023,Evered2025,Bornet2026}.

A central challenge for Rydberg atom based realizations of QSLs as equilibrium states is the requirement of sufficiently low entropies translating to temperatures $T$ significantly below the (pseudo-)spin interaction $J$. For reference, solid-state candidate materials are typically studied in cryostats realizing temperatures in a range $0.01\lesssim T/J\lesssim 0.1$~\cite{Savary2017,Han2012,Wen2019}. Moreover, this temperature is straightforwardly known from cryostat instrumentation. In contrast, for atomic quantum simulators the temperature depends on details of the state preparation protocol and is typically determined only indirectly from measurements of system observables~\cite{Sbierski2024}. As the extracted temperature estimate in general depends on the observable, i.e., the degrees of freedom under consideration, establishing thermalization requires consistency across different observables. Hence, reliable and precise thermometry is essential to quantify the proximity to the low-temperature putative QSL regime and, more generally, to establish a firm theoretical footing of Rydberg atom quantum simulation. 

The challenges for such a thermometry protocol are twofold. First, suitable experimental observables must be identified and measured with high precision. Second, theoretical reference data for these observables must be provided based on the assumed thermal state $\rho\sim \mathrm{exp}({-H/T})$. Here the difficulty lies in the frustrated nature of the models in question, which renders numerical approaches challenging. In particular, quantum Monte-Carlo is hampered by the sign problem~\cite{Sandvik2010}.

We propose a thermometry approach that copes with the above two challenges and demonstrate it using a recent pioneering implementation of a Kagome lattice Rydberg simulator by Bornet et al.~\cite{Bornet2026}. Specifically, we use the local $zz$ spin susceptibility and the nearest-neighbor equal-time correlator as observables, where the former was measured in this platform for the first time (for a different purpose). Second, we provide theoretical estimates of these quantities based on the recently developed dynamic high-temperature expansion (Dyn-HTE) for spin correlations~\cite{BurkardPRB}. The temperature is then estimated by matching experimental values with theoretical predictions for a thermal state at temperature $T$.

Using this approach, we find a temperature $T = 0.55J$ and $s=0.67 \, \mathrm{ln}2$ for the entropy per spin. These relatively large values are in line with the lack of clear QSL signatures~\cite{Bornet2026}. This shows that despite the considerable focus on the (putative QSL) ground state of relevant dipolar XY models in the theoretical literature~\cite{Bintz2024,Machado2026,Mao2026}, the immediate challenge for upcoming experiments rather lies in achieving the order of magnitude reduction of temperature required to faithfully emulate state-of-the-art solid-state experiments \cite{Han2012}. The thermometry protocol proposed here will be valuable to monitor future progress in this direction.

\section{Model and observables}
\begin{figure}
    \centering
    \includegraphics[width=86mm]{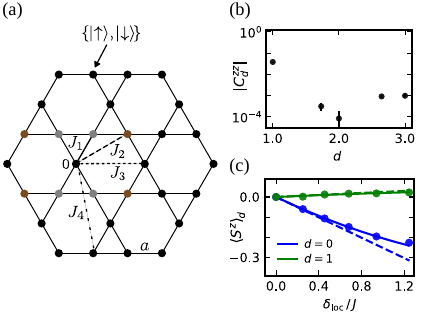}
    \caption{(a) Rydberg atoms (dots) are arranged in a Kagome lattice using optical tweezers. For the correlations computed in this work, we consider an infinite lattice $N\rightarrow \infty$ and truncate the dipolar interaction beyond fourth-nearest neighbors, leading to an effective $J_1-...-J_4$ model. Representative interactions $J_1=J$ (solid), $J_2$ (dashed), $J_3$ (dotted), and $J_4$ (sparsely dotted) are indicated. Sites at distances $d=1$ and $d=\sqrt{3}$ from site $0$ are shown in gray and brown respectively. The number of such sites is $N_1=4=N_{\sqrt{3}}$. (b) Experimentally measured equal-time $zz$-correlators, taken from Fig.~S3 of Ref.~\onlinecite{Bornet2026}. (c) Measured magnetization response $\langle S^z\rangle_d$ to the perturbation $\delta_\text{loc}S_0^z$, from Fig.~5(a) of Ref.~\onlinecite{Bornet2026}. By fitting a nonlinear response ansatz (full lines, cf.~Eq.~\eqref{eq:nonlinear_fit}), we obtain $J\chi^{zz}_0=(-2.63\pm0.03)\cdot10^{-1}$ and $J\chi^{zz}_1=(2.61\pm0.97)\cdot10^{-2}$ for the linear susceptibility (dashed line), for the local and nearest-neighbor case, respectively.}
    \label{fig:exp}
\end{figure}

We first review the experiment by Bornet et al.~\cite{Bornet2026}, which realizes a two-dimensional Kagome lattice with $N=114$ $^{87}\mathrm{Rb}$ atoms trapped in optical tweezers, see Fig.~\ref{fig:exp}(a). Each atom encodes a (pseudo-)spin-$1/2$ in two Rydberg states. The anti-ferromagnetic dipolar XY spin Hamiltonian reads
\begin{align}
    H_\mathrm{XY}&=\frac{J}{2}\sum_{i<j}\frac{a^3}{r_{ij}^3}\left(\sigma^x_i\sigma^x_j+\sigma^y_i\sigma^y_j\right)\label{eq:H}\\
    &=J\sum_{i\neq j}\frac{a^3}{r^3_{ij}}\left(S^x_iS^x_j+S^y_iS^y_j\right)\,,
\end{align}
where $J>0$ denotes the coupling strength ($2\pi \times0.77 \text{MHz}$, we set $\hbar=1$ and $k_B=1$), $a=12\mu \mathrm{m}$ is the nearest-neighbor distance and $r_{ij}=|\mathbf{r}_i-\mathbf{r}_j|$ is the distance between atoms $i$ and $j$. Further, $S_i^{x,y}=\sigma_i^{x,y}/2$ are spin-$1/2$ operators expressed via Pauli matrices $\sigma^{x,y}_i$ at site $i$. In the following, we measure length scales in units of $a$ and energy scales in units of $J$.
A low-energy state of $H_\mathrm{XY}$ is prepared starting from an (approximate) antiferromagnetic product state by using a ramp-down of a staggered $z$-field. The final state shows no signatures of symmetry breaking (e.g.~$\langle S^\alpha_j\rangle=0$ for $\alpha \in \{x,y,z\}$) and is discussed as a candidate for a QSL. To characterize this state two qualitatively different types of observables are measured, see Fig.~\ref{fig:exp}(b,c) for the results reproduced from Ref.~\onlinecite{Bornet2026} which we correct for detection errors \footnote{The measured data is corrected using $\langle S^{z}\rangle_d^{meas}=(1-\epsilon_{\uparrow}-\epsilon_{\downarrow})\langle S^{z}\rangle_d$ for the change of magnetization upon insertion of the local field and $\langle S_{i}^{z}S_{j}^{z}\rangle^{meas}=\left[1-2(\epsilon_{\uparrow}+\epsilon_{\downarrow})\right]\langle S_{i}^{z}S_{j}^{z}\rangle$ for equal-time correlators with $\epsilon_{\downarrow}=0.013$ and $\epsilon_{\uparrow}=0.026$, see Fig.~S2 of Ref.~\onlinecite{Bornet2026}.}.

As a first observable, equal-time $zz$ correlators between a central reference spin (site $i=0$) and all other spins at distance $d=r_{0j}$ are projectively measured and averaged over $N_d$ symmetry-equivalent sites $j$ [see colored sites in Fig.~\ref{fig:exp}(a)],
\begin{equation}
    C^{zz}_d=\frac{1}{N_d}\sum_{|r_{0j}|=d}\langle S^z_0 S^z_j\rangle.
    \label{eq:Cd}
\end{equation}
In the following, we focus on the correlations of nearest-neighbor spins ($d=1$), which are by far the largest in amplitude, see Fig.~\ref{fig:exp}(b). Also note that experimental data for the $xx$-correlator is also available and of similar magnitude \cite{Bornet2026}, however the measurement is complicated by an additional basis rotation (if available at all). The nearest-neighbor equal-time correlator is a standard observable for thermometry in atom-based quantum simulation~\cite{mazurenkoColdatomFermiHubbard2017,Sbierski2024}.

As a second (and novel) observable, distance-resolved static spin susceptibilities of $zz$ type are obtained in the experiment. The original intention was to study spinon Friedel oscillations~\cite{Bornet2026}. For this measurement, a local $z$-field is applied selectively to spin $i=0$, modifying the Hamiltonian according to $H_\mathrm{XY}\to H_\mathrm{XY}+\delta_\text{loc}\,S^z_0$.
The resulting average change in $z$-magnetization $\langle S^z\rangle_d$ of atoms at distance $d=r_{0j}$ is measured as a function of $\delta_\text{loc}$. The distance-resolved linear-response static $zz$-susceptibility is then defined as
\begin{equation}\label{eq:susceptibility}
    \chi^{zz}_{d} = \frac{\mathrm{d}\langle S^z\rangle_d}{\mathrm{d}\delta_\text{loc}}\,\big|_{\delta_\text{loc}=0}\,=-\int_0^{1/T}\!\!\!\mathrm{d}\tau\,\left\langle e^{H\tau}S_{0}^{z}e^{-H\tau}S^z_j\right\rangle,
\end{equation}
where last equality indicates the theoretical linear-response expression based on the imaginary-time evolved correlator \cite{bruusManyBodyQuantum2004}. In the following, we focus on the \emph{local} $(d=0)$ susceptibility (which has the largest amplitude \cite{Bornet2026}), for which a HTE based on Eq.~\eqref{eq:susceptibility} yields
\begin{equation}
    \chi^{zz}_0=-\frac{1}{4T}+O\left(\frac{J}{T^2}\right)\,.
\end{equation}
This local susceptibility has certain conceptual features which render it suitable and complementary to the nearest-neighbor equal-time correlator in the context of thermometry. (i) According to the first equality in Eq.~\eqref{eq:susceptibility}, it is based on the measurement of a single-site expectation value, which can be less susceptible to noise compared to the two-site observable in \eqref{eq:Cd}. (ii) It is a purely \emph{local} observable whereas the \emph{local} correlator $C_{d=0}^{\alpha \alpha}=1/4$ always evaluates to a constant and thus contains no information about the state. (iii) It can be measured with high precision: By ramping the local field $\delta_\text{loc}$ up to values of order $O(J)$, one can obtain $\chi^{zz}_0$ by fitting the corresponding values of $\langle S^z\rangle_0(\delta_\text{loc}) \sim O(J/T)$ with a non-linear response form,
\begin{equation}
       \langle S^z\rangle_d(\delta_\text{loc})=\chi^{zz}_d\,\delta_\text{loc}+\chi_d^{zz(2)} \,\frac{\delta_\text{loc}^2}{2}\,, \label{eq:nonlinear_fit}
\end{equation}
(written for general $d$), see full lines in Fig.~\ref{fig:exp}(c) for $d=0,1$. Discarding the 2nd-order susceptibility $\chi_d^{zz(2)}$ (which is slightly harder to compute in theory), we obtain an accurate estimate for $\chi_d^{zz}$, see caption of Fig.~\ref{fig:exp} for the values. These estimates are robust with respect to the upper end of the $\delta_\text{loc}$ fitting range: adding or removing one data point at large $\delta_\text{loc}$ affects the fitted parameters only within their uncertainties.


\section{Thermometry}\label{sec:thermometry}
We now address the second challenge in thermometry, i.e.~the comparison of the measured data to theoretical results obtained for the assumed thermal state and determination of the matching temperature. We obtain consistent temperature estimates from the nearest-neighbor equal-time spin correlator $C_1^{zz}$, the local susceptibility $\chi_0^{zz}$, and the (significantly weaker) nearest-neighbor susceptibility $\chi_1^{zz}$, which validates the consistency of the assumed thermal state across observables.

To obtain theoretical predictions, quantum Monte-Carlo simulations \cite{Sandvik2010,sadouneEfficientScalablePath2022} are often the method of choice. However, this approach is hindered by the frustrated nature of the anti-ferromagnetic Kagome lattice Hamiltonian \eqref{eq:H} and an alternative method is required. Except at very low temperatures, high-order HTE~\cite{Oitmaa2006} provides an established, frustration-insensitive approach for computing distance-resolved equal-time correlators $C_d^{zz}$~\cite{hehnHightemperatureSeries2017}. To also simulate the static susceptibilities $\chi_d^{zz}$, a dynamic variant of the the HTE is required to deal with the imaginary-time evolved correlator on the right-hand side of Eq.~\eqref{eq:susceptibility}. This has been achieved recently (with the involvement of some of us) in Ref.~\onlinecite{BurkardPRB} for $zz$-correlators of Heisenberg spin models with a single coupling constant ($J_1$) and we here adapt this Dyn-HTE method to the present case of spin-$1/2$ XY models. 
We truncate the interaction range and only keep the exchange coupling for $r_{ij}\leq r_{max}=\sqrt{7}=2.65$ corresponding to fourth-nearest neighbors. We hence work with an effective $J_1-...-J_4$ model where $J_1=J$, see Fig.~\ref{fig:exp}(a). We reach an expansion of 8th order in $J/T$ (the numerical bottleneck is the graph-embedding step \cite{BurkardPRB} in the highly connected lattice graph). As shown in App.~\ref{app:truncation}, this interaction truncation is well converged for the intermediate temperature ranges currently relevant to experiment. Our results are reported for the simulation in the thermodynamic limit $N\rightarrow \infty$, a comparison with a simulation at the experimental system size $N=114$ revealed no detectable finite-size effects at the relevant $J/T$. Finally, note that both $\chi_d^{zz}$ and $C_d^{zz}$ can be calculated from Dyn-HTE. The former is identified with the Matsubara correlator at vanishing frequency while the latter is obtained after summation over Matsubara frequencies.
  
\begin{figure}
    \centering
    \includegraphics[width=0.9\linewidth]{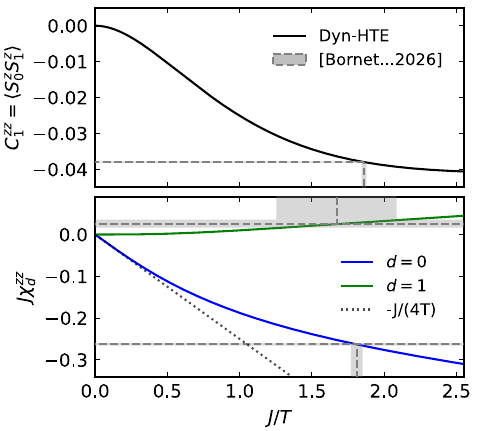}
    \caption{Top: Nearest-neighbor equal-time $zz$-correlator $C^{zz}_1$ as a function of $J/T$. By comparison with the experimental value from panel Fig.~\ref{fig:exp}(b), shown as a horizontal dashed line, a corresponding temperature of $J/T=1.85(2)$ can be estimated. Gray regions indicate error bounds. Bottom: Distance resolved static $zz$-susceptibilities $\chi^{zz}_d$ as a function of $J/T$. The experimental linear response susceptibilities $\chi^{zz}_d$ from Fig.~\ref{fig:exp}(c) (horizontal dashed lines) correspond to $J/T=1.81(4)$ and $J/T=1.7(4)$, respectively. The theoretical Dyn-HTE data shown in both panels represent the [4,4] Padé approximants which agree well with the [3,5], [3,4], and [3,3] approximants (not shown) and can thus be considered accurate \cite{Oitmaa2006,BurkardPRB}.}
    \label{fig:Dyn-HTE}
\end{figure}

Our (Dyn-)HTE results for the observables versus inverse temperature $J/T$ are shown in Fig.~\ref{fig:Dyn-HTE} as full lines. We start with the nearest-neighbor correlator $C_1^{zz}$ shown in the top panel which we match to the experimentally determined value from Fig.~\ref{fig:exp}(b), see horizontal dashed line (with gray error-bar). We obtain $J/T=1.85(2)$, see the vertical bar. Finally, the bottom panel of Fig.~\ref{fig:Dyn-HTE} provides an analogous analysis for the local and nearest-neighbor susceptibility $\chi_d^{zz}$ and yields temperature estimates $J/T = 1.81(4)$ (for $d=0$) and $J/T=1.7(4)$ (for $d=1$). As anticipated, all obtained temperatures are mutually consistent within error bars, providing evidence for a thermal state. We skip a similar analysis for equal-time correlators at larger distance $d>1$ as they are orders of magnitude smaller with much larger relative errors, see Fig.~\ref{fig:exp}(b).

\section{Entropy estimate}
Besides temperature $T$, the entropy per spin $s$ is an important quantity in Rydberg tweezer experiments \cite{Sbierski2024,homeierSupersolidityRydbergTweezer2025,Bornet2026}. The entropy is lower-bounded by the entropy of the initially prepared product state, which is straightforward to measure via the magnetization. In the Kagome lattice experiment of Ref.~\onlinecite{Bornet2026} this initial entropy per spin $s=S/N$ is $s_{t=0} = 0.3\ln 2$. During the subsequent ramp-down of staggered fields towards the final correlated state, diabatic and decoherence effects will increase the entropy.

\begin{figure}
    \centering
    \includegraphics[width=0.9\linewidth]{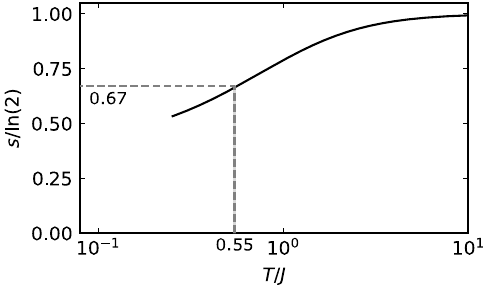}
    \caption{Entropy per spin $s$ of the dipolar XY model \eqref{eq:H} on the infinite Kagome lattice obtained from an order six HTE of the free energy with well-converged interaction cutoff range $r_{max}=10$ [cf.~App.~\ref{app:entropy}]. Here data based on a stable [3,3] Padé approximant is shown. The dashed lines indicate the temperature obtained from thermometry in Fig.~\ref{fig:Dyn-HTE} and the associated entropy. Gray regions indicate error bounds.}
    \label{fig:entropy_per_spin}
\end{figure}

We provide an estimate of the final entropy in the Kagome lattice experiment \cite{Bornet2026} based on the assumption that the state is thermal with the temperature determined in the previous section. To relate temperature to entropy, we use $s=-\beta\partial_\beta w + w$ with $w=\mathrm{ln}(Z)/N$ the dimensionless free energy per spin and $Z$ the partition function. We again employ a HTE (for $w$) and work with a cumulant expansion \cite{Oitmaa2006} building on free-graph diagrammatics. We reach expansion order six in $J/T$. The free graphs allow for a very large interaction truncation range $r_{max}=10$ in the approximation of lattice sums which is well converged. We refer to App.~\ref{app:entropy} for further details. The resulting temperature dependence of the entropy is shown in Fig.~\ref{fig:entropy_per_spin}. For $J/T=1.83(3)$ determined as the mean of the temperatures found above from $C_1^{zz}$ and $\chi^{zz}_0$, we obtain an entropy per spin $s=0.67(1)\ln 2$. This indicates more than a doubling of the initial state entropy during the ramp-down of the staggered field. For a truly adiabatic (infinitely slow) ramp, the initial entropy $s_{t=0}=0.3\ln2$ would be maintained and would correspond to a significantly lower temperature $J/T=O(10)$ (a more precise estimate requires HTE to higher orders).

Our obtained entropy significantly exceeds the estimate $s \simeq 0.6\ln 2$ reported in Ref.~\onlinecite{Bornet2026} which is based on a protocol where the preparation ramp is followed immediately by a reverse ramp. The final staggered magnetization is measured and found to correspond to an entropy $s_{t=2t_{\mathrm{ramp}}} = 0.9\ln 2$ close to an infinite temperature state with $s_\infty=\ln 2$. If entropy production was symmetric between both ramps, the estimate $s \simeq 0.6\ln 2$ would follow~\cite{Bornet2026}. Our more reliable thermometry approach suggests that this symmetry assumption is violated.

\section{Conclusion}
We demonstrate a thermometry protocol applied to a recent Kagome lattice dipolar Rydberg quantum simulator experiment~\cite{Bornet2026} based on two conceptually different observables, the nearest-neighbor equal-time correlations and the static local susceptibility. The temperatures extracted from these observables are in excellent mutual agreement and support the fundamental underlying assumption of a thermal state. Our final estimate for the temperature is $J/T = 1.83(3)$. Our findings suggest that improvements on the preparation ramp protocol might be needed to reach the strongly correlated low temperature regime $J/T\gtrsim10$ usually studied in solid-state samples. For the entropy per spin we obtain $s=0.67(1)\ln 2$, which is significantly larger than the estimate reported in Ref.~\onlinecite{Bornet2026}.

In the past, a similar thermometry analysis \cite{Sbierski2024} (based only on equal-time correlators) was conducted for a dipolar XY square lattice quantum simulator experiment \cite{chenContinuousSymmetryBreaking2023}. To treat the frustrated case of anti-ferromagnetic interactions theoretically \cite{Sbierski2024}, a pseudo-Majorana functional renormalization group method \cite{Niggemann2021,mullerPseudofermionFunctionalRenormalization2024} was employed. The Dyn-HTE method used here is equally oblivious to 
the model's sign problem and long-range nature of the interactions, while offering appealing features such as a straightforward error-control at reduced temperatures.

Finally, we propose to use the protocol presented here as a blueprint for future thermometry applications in cold-atom quantum simulation, and in particular for Rydberg tweezer arrays. For the theory side, the Dyn-HTE method is available as an efficient and versatile open-source software package which can be straightforwardly adapted to the desired lattice geometry \cite{Dyn-HTEsoftware_v1_0, zenodoKagomeThermometry}.

\section*{DATA AVAILABILITY}
The data and program codes that support the findings of this article are openly available \cite{zenodoKagomeThermometry}.

\begin{acknowledgments}
We thank 
Christian Groß, Lode Pollet and Guillaume Bornet for useful discussions and Andreas Alexander Buchheit for providing the quasi-exact lattice sums in the rightmost column of Tab.~\ref{tab:coefficients}.
We acknowledge computational support by the state of Baden-Württemberg through bwHPC and the German Research Foundation (DFG) through Grant No.~INST 40/575-1 FUGG (JUSTUS 2 cluster).
We acknowledge funding from the Deutsche
Forschungsgemeinschaft (DFG, German Research Foundation) through the Research Unit FOR 5413 (project N2), Grant No.~465199066.
\end{acknowledgments}

\appendix

\begin{widetext}


\section{Effects of truncation on interaction range}\label{app:truncation}

\begin{figure}
    \centering
    \includegraphics[width=0.45\linewidth]{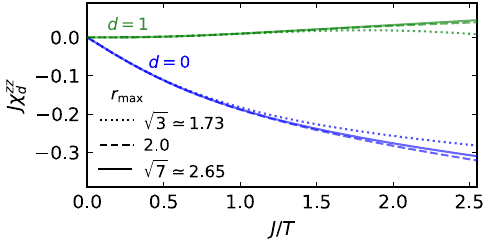}
    \caption{Local (blue) and nearest-neighbor (green) static $zz$ susceptibilities computed with Dyn-HTE for different truncation ranges ($r_\mathrm{max}=\sqrt{3},2,\sqrt{7}$), showing good convergence.}
    \label{fig:truncation}
\end{figure}
Our theoretical calculations throughout this work depend on an chosen truncation range $r_\mathrm{max}$ (in units of nearest-neighbor distance $a$) for the dipolar interaction in the Hamiltonian \eqref{eq:H}. To ensure that our results are converged in $r_{max}$, we here systematically examine different truncation ranges for the observables considered. In Fig.~\ref{fig:truncation}, we compare static susceptibility results for $r_{max}\in \{ \sqrt{3},2,\sqrt{7} \}$ and find only minor differences between the largest two $r_{max}$. This justifies our choice to truncate the dipolar interaction beyond fourth nearest neighbors ($r_\mathrm{max}=\sqrt{7}$). The same conclusion holds for the equal-time correlator which is obtained by a sum of the Matsubara correlators over Matsubara frequencies \cite{BurkardPRB} (the static susceptibility corresponds to the Matsubara correlator at $i\omega_n=0$).

For the entropy calculation detailed in App.~\ref{app:entropy}, the usage of \emph{free} graphs \cite{Oitmaa2006} simplifies lattice sums so that a very large value of $r_{max}=10$ was reached. The lattice sums $\phi_{(nx)}$ calculated for $r_{max}=10$ provided in Tab.~\ref{tab:coefficients} agree to sub-percent accuracy with the quasi-exact results obtained by the recently proposed zeta-function method~\cite{Buchheit2024,buchheitEpsteinZetaMethod2025,robles-navarroExactLatticeSummations2025} (see rightmost column), at least for the set of graphs where the latter is available.


\section{Entropy from HTE of free energy}\label{app:entropy}

\noindent We here provide details of the entropy calculation which proceeds via a HTE of the free energy. We define the inverse temperature $\beta=1/T$ and rewrite the Hamiltonian
\begin{equation}
    -\beta H = -\beta J\sum_{i\neq j}\frac{1}{r_{ij}^3}\left(S^x_iS^x_j+S^y_iS^y_j\right)=\frac{1}{2}\sum_{i,j}\sum_{\gamma,\gamma^\prime} v(ij,\gamma\gamma^\prime)S^\gamma_iS^{\gamma^\prime}_j\,,
\end{equation}
where we define $v(ij,\gamma\gamma^\prime)\equiv-\beta J r^{-3}_{ij}(1-\delta_{ij})(1-\delta_{\gamma\gamma^\prime})$ and introduce the flavor variable $\gamma\in\{+,-\}$ and $S^\pm\equiv S^x\pm iS^y$. Moreover, let 
\begin{equation}
    Z\equiv\mathrm{Tr}\left[e^{-\beta H}\right]\,,\qquad
    \rho\equiv\frac{1}{Z}e^{-\beta H}\,,\qquad
    S\equiv-\mathrm{Tr}\left[\rho\ln\rho\right]\,,
\end{equation}
denote the partition function $Z$, the thermal state $\rho$, and the von Neumann entropy $S$. Using $\ln\rho = -\beta H - \ln Z$, we obtain
\begin{equation}
    S = \beta \langle H\rangle + \ln Z\,,
\end{equation}
with $\langle H\rangle\equiv\mathrm{Tr}[\rho H]$. Since
\begin{equation}
    \partial_\beta \ln Z = \frac{1}{Z}\mathrm{Tr}\left[\partial_\beta e^{-\beta H}\right] = -\mathrm{Tr}[\rho H] = -\langle H\rangle\,,
\end{equation}
we can express the entropy per spin $s\equiv S/N$ solely in terms of the partition function,
\begin{equation}\label{eq:entropy_free_energy}
    s(\beta)=\frac{1}{N}\left(-\beta\partial_\beta\ln Z + \ln Z\right)=\frac{1}{N}\left(-\beta\partial_\beta W + W\right)\,,
\end{equation}
where $W\equiv\ln Z$ is termed the \emph{dimensionless free energy}. 

To obtain a HTE of $W$, we expand the partition function as 
\begin{align}
    Z&=\sum_{n=0}^\infty \frac{1}{2^n n!}\,\mathrm{Tr}\left[ \left(\sum_{i,j}\sum_{\gamma,\gamma^\prime} v(ij,\gamma\gamma^\prime)S^\gamma_iS^{\gamma^\prime}_j\right)^n\,\right]\nonumber\\[1em]
    &=2^N\sum_{n=0}^\infty \frac{1}{2^n n!}\left(\sum_{i_1,\dots,i_{2n}}\sum_{\gamma_1,\dots,\gamma_{2n}} v(i_1i_2,\gamma_1\gamma_2)\cdots v(i_{2n-1}i_{2n},\gamma_{2n-1}\gamma_{2n})\langle S^{\gamma_1}_{i_1}\cdots S^{\gamma_{2n}}_{i_{2n}}\rangle_0\right)\,,
\end{align}
where $\langle\cdots\rangle_0=2^{-N}\mathrm{Tr}[\cdots]$ refers to the expectation value in the $v=0$ ensemble, i.e.~free spins or infinite temperature. Since $Z$ is the moment generating function, we can expand the free energy $W=\ln Z$ in terms of cumulants
\begin{equation}
    W=N\ln 2 + \sum_{n=0}^\infty \frac{1}{2^n n!}\left(\sum_{i_1,\dots,i_{2n}}\sum_{\gamma_1,\dots,\gamma_{2n}} v(i_1i_2,\gamma_1\gamma_2)\cdots v(i_{2n-1}i_{2n},\gamma_{2n-1}\gamma_{2n})\langle S^{\gamma_1}_{i_1}\cdots S^{\gamma_{2n}}_{i_{2n}}\rangle_{0,c}\right)\,,\label{eq:HTE_free_energy}
\end{equation}
where the cumulants are defined recursively
\begin{equation}
    \langle\prod_{n=1}^k A_n\rangle_c\equiv\langle\prod_{n=1}^k A_n\rangle-\sum_\mathcal{P}\prod_{B\in\mathcal{P}}\langle\prod_{n\in B}A_n\rangle_c\,,
\end{equation}
with $\mathcal{P}$ denoting the set of all nontrivial partitions of $\{1,\dots,k\}$~\cite{Oitmaa2006,Domb1974,Schneider2024}. Eventually,
Eq.~\eqref{eq:HTE_free_energy} admits a diagrammatic representation in terms of connected, topologically distinct \emph{free graphs}. The graphs which are relevant for the free energy of the XY model are shown in Fig.~\ref{fig:graphs6}. Each graph (a multi-graph with positive integer edge weights) is assigned a contribution according to the following rules~\cite{Domb1974}:
\begin{enumerate}
    \item Assign a position label $i$ to each vertex.
    \item For each edge joining vertices $i$ and $j$ write a factor $v(ij,\gamma\gamma^\prime)$.
    \item For each vertex $i$ connected to $l$ edges write a factor $\langle S^{\gamma_1}_i\cdots S^{\gamma_l}_i\rangle_{0,c}$.
    \item Sum over all flavors $\gamma\in\{\pm\}$.
    \item Sum over all permutations of edges.
    \item Sum each vertex label $i$ freely over the entire lattice.
    \item Divide by the symmetry factor of the graph (number of graph automorphisms respecting the edge weights times the product of factorials of the edge weights).
\end{enumerate}

\noindent Let us label all topologically distinct connected graphs at order $n$ with an additional index $x=a,b,c,\dots$, and the corresponding contributions to the free energy according to rules 1 to 5 by $w_{(nx)}$. Since the arguments $i_1,\dots,i_{2n}$ are summed, graphs in which these arguments have been permuted in a way that leaves the topology unchanged are considered identical. The number of such permutations is $2^n n! / g_{(nx)}$, where $g_{(nx)}$ is the symmetry factor mentioned in rule 7. In the classical (commuting) case, all orderings of the interaction operators are equivalent, and each graph contributes only once. In contrast, in the non-commuting case different orderings of the same set of interaction operators correspond to distinct operator strings. Therefore, one must explicitly sum over all permutations of the interaction operators (rule 5). Using this diagrammatic representation, the dimensionless free energy per spin $w\equiv W/N$ can be expressed as
\begin{equation}
    w=\ln 2 + \sum_{n=1}^\infty\sum_x \frac{w_{(nx)}\phi_{(nx)}}{g_{(nx)}}\,,\label{eq:HTE_free_energy_graphs}
\end{equation}
where we acquire the contribution per spin, by fixing one vertex label of each graph and summing the remaining vertex labels freely. This lattice sum is denoted by $\phi_{(nx)}$ and is easily computed numerically using linear algebra expressions based on the (truncated) matrix $J_{ij}$ with entries $1/r^3_{ij}$ and zero for $r_{ij}>r_{max}$ and on the diagonal.\\
All contributing graphs up to order $n\leq 6$ are shown in Fig.~\ref{fig:graphs6} where by $\mathrm{U}(1)$ spin-rotation symmetry, only graphs with an even number of incident edges at each vertex can contribute~\cite{Schneider2024} and an equal number of $+$ and $-$ flavors per vertex is required.
\begin{figure}
    \centering
    \includegraphics[width=0.8\linewidth]{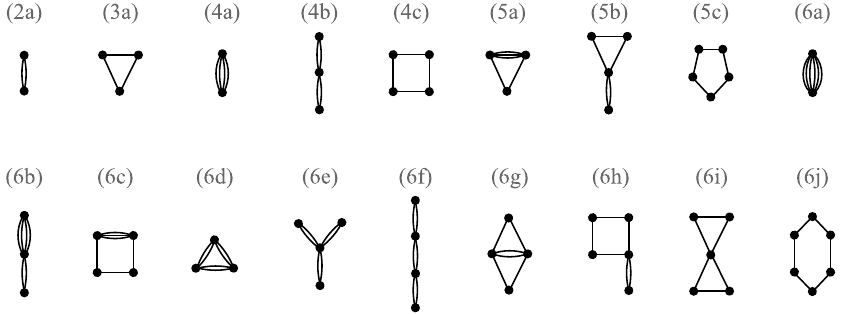}
    \caption{All topologically distinct connected graphs with order $n\leq6$ that yield a nonvanishing contribution to Eq.~\eqref{eq:HTE_free_energy}.}
    \label{fig:graphs6}
\end{figure}
Combining these graphs with Eq.~\eqref{eq:HTE_free_energy_graphs}, the rules above, and Tab.~\ref{tab:coefficients}, we obtain \begin{align}\label{eq:free_energy_order6}
    w=&\ln 2 + \frac{(\beta J)^2}{8}\phi_{(2a)} - \frac{(\beta J)^3}{24}\phi_{(3a)} + (\beta J)^4\left(\frac{\phi_{(4a)}}{48}-\frac{\phi_{(4b)}}{24}+\frac{\phi_{(4c)}}{64}\right)-(\beta J)^5\left(\frac{\phi_{(5a)}}{32}-\frac{\phi_{(5b)}}{24}+\frac{\phi_{(5c)}}{160}\right)\nonumber\\[1em]
    &+(\beta J)^6\left(\frac{17\phi_{(6a)}}{1440}-\frac{19\phi_{(6b)}}{480}+\frac{\phi_{(6c)}}{64}-\frac{23\phi_{(6d)}}{2880}+\frac{\phi_{(6e)}}{60}+\frac{9\phi_{(6f)}}{640}+\frac{41\phi_{(6g)}}{1920}-\frac{\phi_{(6h)}}{48}-\frac{\phi_{(6i)}}{96}+\frac{\phi_{(6j)}}{384}\right)\nonumber\\[1em]
    &+ O(\beta^7)\,.
\end{align}
Inserting the lattice sums for the truncated dipolar interaction on the Kagome lattice [cf.~Tab.~\ref{tab:coefficients}] yields
\begin{equation}
    w(\beta)\simeq\ln(2) + 0.54\cdot (\beta J)^2 - 0.43\cdot (\beta J)^3 + 0.34\cdot (\beta J)^4 - 0.40\cdot (\beta J)^5 + 0.76\cdot (\beta J)^6 + O(\beta^7)\,,
\end{equation}
and thus finally, using Eq.~\eqref{eq:entropy_free_energy},
\begin{equation}
    s(\beta)=w-\beta\partial_\beta w\simeq\ln(2) - 0.54\cdot (\beta J)^2 + 0.86\cdot (\beta J)^3 - 1.01\cdot (\beta J)^4 + 1.59\cdot (\beta J)^5 - 3.80\cdot (\beta J)^6 + O(\beta^7)\,,
\end{equation}
the [3,3] Padé approximant of which is shown in Fig.~\ref{fig:entropy_per_spin}. We have benchmarked this approach versus exact diagonalization for an all-to-all connected toy model on $N=6$ sites. 
\begin{table}[t]
    \centering
    \begin{tabular}{c c c c c}
        \toprule
        Graph & $g_{(nx)}$ & $w_{(nx)}$ & \makecell{$\phi_{(nx)}$ \\ ($r_{max}=10$)} & \makecell{$\phi_{(nx)}$ \\ (full)}\\
        \midrule
        
        (2a)   & $4$ & $1/2$ & $4.284$ & $4.283$\\
        
        (3a)   & $6$ & $1/4$ & $10.275$ & $10.2786$\\
        
        (4a)   & $48$ & $1$ & $4.007$ & $4.0070$\\
        (4b)   & $8$ & $-1/3$ & $18.35$ & $18.3509$\\
        (4c)   & $8$ & $1/8$ & $65.044$ & $65.1064$\\
        
        (5a)   & $12$ & $3/8$ & $7.328$ & $7.3283$\\
        (5b)   & $4$ & $-1/6$ & $44.014$ & $44.0316$\\
        (5c)   & $10$ & $1/16$ & $320.517$ & $321.3446$\\
        
        (6a)   & $1440$ & $17$ & $4.000$ & $4.0002$\\
        (6b)   & $48$ & $-19/10$ & $17.165$ &$17.1653$\\
        (6c)   & $12$ & $3/16$ & $48.776$&$48.7922$\\
        (6d)   & $48$ & $-23/60$ & $4.666$&$4.6658$\\
        (6e)   & $48$ & $4/5$ & $78.604$&$78.6115$\\
        (6f)   & $16$ & $9/40$ & $78.604$&$78.6115$\\
        (6g)   & $8$ & $41/240$ & $13.847$\\
        (6h)   & $4$ & $-1/12$ & $278.627$&$278.9029$\\
        (6i)   & $8$ & $-1/12$ & $105.573$&$105.6508$\\
        (6j)   & $12$ & $1/32$ & $1866.744$ & $1876.3773\dots$\\     
        \bottomrule
    \end{tabular}
    \caption{Graph evaluations for HTE of the free energy according to Eq.~\eqref{eq:HTE_free_energy_graphs} with symmetry factor $g_{(nx)}$, flavor sum $w_{(nx)}$ and lattice sum $\phi_{(nx)}$ for the truncated and full dipolar interaction on the Kagome lattice. The full lattice sums are available for all but one graph using a recently established method \cite{Buchheit2024,buchheitEpsteinZetaMethod2025,robles-navarroExactLatticeSummations2025}}
    \label{tab:coefficients}
\end{table}

\end{widetext}

\bibliography{references}

\end{document}